\documentclass[useAMS,usenatbib]{mn2e}
\usepackage{graphics, graphicx}
\newif\ifAMStwofonts

\def\lapp{\ifmmode\stackrel{<}{_{\sim}}\else$\stackrel{<}{_{\sim}}$\fi}
\def\gapp{\ifmmode\stackrel{>}{_{\sim}}\else$\stackrel{>}{_{\sim}}$\fi}

\newcommand{\degrees}[1]{\ensuremath{#1^\circ}}

\title[Mapping the magnetosphere of PSR B1055--52]
{Mapping the magnetosphere of PSR B1055--52}
\author[Weltevrede \& Wright]
{Patrick Weltevrede$^1$\thanks{E-mail: Patrick.Weltevrede@atnf.csiro.au} and Geoff Wright$^{2,1}$\\
$^{1}$Australia Telescope National Facility, CSIRO, P.O. Box 76, Epping, NSW 1710, Australia.\\
$^2$Astronomy Centre, University of Sussex, Falmer, BN1 9QJ, UK
}
\date{}
\pagerange{\pageref{firstpage}--\pageref{lastpage}}
\pubyear{2008}
\begin{document}
\maketitle
\label{firstpage}

\begin{abstract}
We present a geometric study of the radio and $\gamma$-ray pulsar
B1055--52 based on recent observations at the Parkes radio
telescope. We conclude that the pulsar's magnetic axis is inclined at
an angle of $\degrees{75}$ to its rotation axis and that both its
radio main pulse and interpulse are emitted at the same height above
their respective poles. This height is
unlikely to be higher or much lower than 700 km, a typical value for
radio pulsars. 

It is argued that the radio interpulse arises from emission formed on
open fieldlines close to the magnetic axis which do not pass through
the magnetosphere's null (zero-charge) surface. However the main pulse
emission must originate from fieldlines lying well outside the polar
cap boundary beyond the null surface, and farther away from the
magnetic axis than those of the outergap region where the single
$\gamma$-ray peak is generated.  This casts doubt on the common
assumption that all pulsars have closed, quiescent, corotating regions
stretching to the light cylinder.
\end{abstract}


\begin{figure*}
\includegraphics[height=0.99\hsize,angle=270]{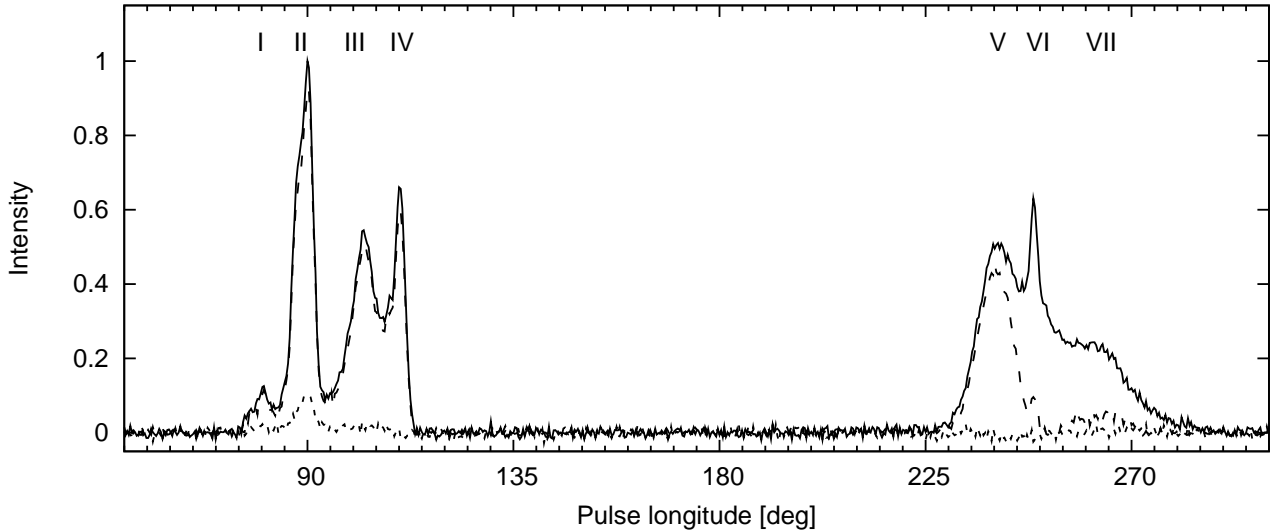}
\caption{\label{Fig1055profiles20}The pulse profile of the 1369~MHz
timing data of PSR~B1055--52 at 20~cm showing total intensity (solid
line), linear polarization (dashed line) and circular polarization
(dotted line).}
\end{figure*}

\section{Introduction}

PSR~B1055--52 is a middle-aged (0.5 Myr) energetic
($\dot{E}=3\times10^{34}$~erg~s$^{-1}$) pulsar which has been detected
as a source of pulsed $\gamma$-rays. In their study of energetic
pulsars, \cite{wj08b} singled out PSR~B1055--52 for comment since it
is exceptional in having both an interpulse and wide radio components,
with the interpulse (IP) well separated from the main pulse (MP).  In
general the presence of an IP is a strong clue to a pulsar's geometry,
and suggests that a study of the radio emission of PSR~B1055--52 may
give insight into the structure of the magnetosphere and the origin of
its $\gamma$-rays.

PSR~B1055--52 was discovered with the Molongolo telescope by
\cite{vl72}, who noted a possible IP in one of their records. The IP
was confirmed by \cite{mhak76}, who commented that the pulse profile
resembles that of the Crab pulsar. \cite{ran83} noted that the MP-IP
separation shows no variation with frequency, which is additional
evidence that the pulsar's magnetic axis is highly inclined to the
rotation axis.  \cite{lm88} concluded that the observed polarisation
position angle (PA) swings of the MP and IP suggest an angle of
inclination of the magnetic axis of $\degrees{75}$, without ruling out
higher values around $\degrees{90}$. This general view was supported
in a complex study published by \cite{big90a}.

PSR~B1055--52 is relatively close, which makes it possible to observe
its emission over a wide range of wavelengths. The pulsar's $DM$
distance is only $0.72\pm0.2$~kpc \citep{cl02}, roughly consistent
with the X-ray data (e.g. \citealt{of93}) and the non-thermal radio
source around PSR~B1055--52 (\citealt{cra97}).

The EGRET detector (Energetic Gamma-Ray Experiment Telescope) on board
of the {\em Compton Gamma Ray Observatory} (CGRO) revealed that
PSR~B1055--52 is a pulsed $\gamma$-ray source
(\citealt{tbb+99}). Unlike most $\gamma$-ray profiles (but in common
with PSR~B1706--44), that of PSR~B1055--52 does not show two narrow
peaks separated by 0.4--0.5 in phase, but has a broad $\gamma$-ray
profile with possible sub-peaks at the leading and trailing edge of
the profile which peak 0.25 and 0.05 in phase before the radio MP
\citep{tbb+99}. The leading side of the profile has also been seen
with the COMPTEL detector (0.75--30 MeV), while only the trailing side
of the profile is seen above 2 GeV by EGRET, suggesting a longitude
dependent spectral index.

Another feature which makes PSR~B1055--52 an interesting source at
high energies is its spin-down age of 0.5 Myr, making PSR~B1055--52
(at the time of writing) the oldest of the known $\gamma$-ray
pulsars. Its period of 0.197 seconds makes it also the slowest
(excluding Geminga).  This status as being both an `old' young pulsar
and, at the same time a `young' old pulsar (most radio pulsars are
over 1 Myr old) suggests that it may possess characteristics of both
and give insight into the link between high-energy production in the
outer magnetosphere and radio emission above the polar cap. It has
famously been dubbed as one of the `Three Musketeers' \citep{bt97},
along with PSR~B0656+14 and Geminga, because these pulsars were found
to emit solely thermal X-rays which are interpreted as the product of
initial neutron star cooling.

The geometry linking the high-energy and radio profiles was recently
studied by \cite{wqxl06}, who made use of the PA analysis of
\citealt{lm88}. They concluded that that the radio MP and IP are
emitted from separate poles with the $\gamma$-ray emission coming from
the same pole as the radio MP. They claim that the IP and MP come from
different heights, but do not consider the effect of retardation and
aberration on the PA-curve. In their model the radio emission occurs
on fieldlines which are confined to a polar cap bounded by the last
closed fieldlines which, in a dipole geometry, touch the light
cylinder. Here we use new data to re-analyse the PA-swings and to
create a geometry which is fully consistent with retardation and
aberrational effects. A full understanding of the geometry is an
important prelude to undertaking an analysis of the pulsar's complex
single pulse behaviour, which will appear in a subsequent paper.

The organisation of this paper is as follows: in Sect. \ref{SectObs}
details about the radio data used are provided, in
Sect. \ref{SectProfile} the properties of the pulse profile of
PSR~B1055--52 are described and its most likely geometry is derived in
Sect. \ref{SectGeometry}. In Sect. \ref{SectBeamIllumination} we
identify the active fieldlines of the MP and IP and the implications
of our results are discussed in Sect. \ref{SectDiscussion}.

\section{Observations}
\label{SectObs}

The pulse profile of PSR~B1055--52 was obtained at three different
frequencies by the Parkes 64-m radio telescope located in
Australia. During an observing session carried out from 2006 August 24
to 27 data was recorded using the H-OH receiver (1369~MHz centre
frequency, 256~MHz bandwidth split into 1024 frequency channels with
an equivalent system flux density of 43~Jy; \citealt{joh02}) and the
10/50~cm receiver (3094/653~MHz centre frequency, 1024/64~MHz
bandwidth split into 1024/512 frequency channels with an equivalent
system flux density of 49/57~Jy on a cold sky). The signals from the
two linear feeds of the receiver were converted into Stokes
parameters, resampled and folded at the pulse period by a digital
filterbank using 512 bins across the profile. In addition we also used
the pulse profile with a slightly higher time resolution (1024 bins
across the profile) obtained using the 20~cm multibeam receiver
(1369~MHz centre frequency, 256~MHz bandwidth split into 1024
frequency channels with an equivalent system flux density of 35~Jy on
a cold sky) for the Fermi timing program. This program started in
April 2007 with the aim to provide timing solutions which can be used
to fold $\gamma$-ray data obtained by the Fermi satellite
\citep{sgc+08}. The profile is the sum of 33 individual observations
with a total length of 6.6 hours. For details about the
data-processing we refer to \cite{wj08b}.

\section{Pulse profile}
\label{SectProfile}

\subsection{Profile morphology}

The shape of the pulse profile of PSR~B1055--52 is quite complex, as
one can see in Fig. \ref{Fig1055profiles20}. In a similar way to
\cite{big90a} (who did not detect the first component of the MP), we
call the four most prominent peaks of the MP peaks I to IV and the
three peaks in the IP V to VII.  Long integrations are required to
obtain stable profiles, with evidence for persistent intensity changes
in component III (also noted by \citealt{big90a}).  The peak-to-peak
separation between the MP and IP is $\sim\degrees{159}$, well below
$\degrees{180}$, so it is not immediately obvious that we are dealing
with emission from both poles of the neutron star.

As one can see in Fig. \ref{Fig1055profiles20}, the profile of the MP
is highly linearly polarized. This is also the case for the leading
edge of the IP, but at later pulse longitudes it becomes almost completely
depolarized (e.g.  \citealt{mhma78}). Notice that all three components
of the IP have a distinct linear polarization peak, although they are
much weaker for components VI and VII (Fig. \ref{Fig1055profiles20}).
The profiles of the degree of linear polarization at 10~cm and 50~cm
(not shown) look very similar to that at 20~cm.

Component II of the MP has a significant amount of circular
polarization, although it is weak. There is a hint that the trailing
part IP also has a significant amount of circular polarization, but
with an opposite sign.

\subsection{Frequency evolution of the pulse profile}

\begin{figure}
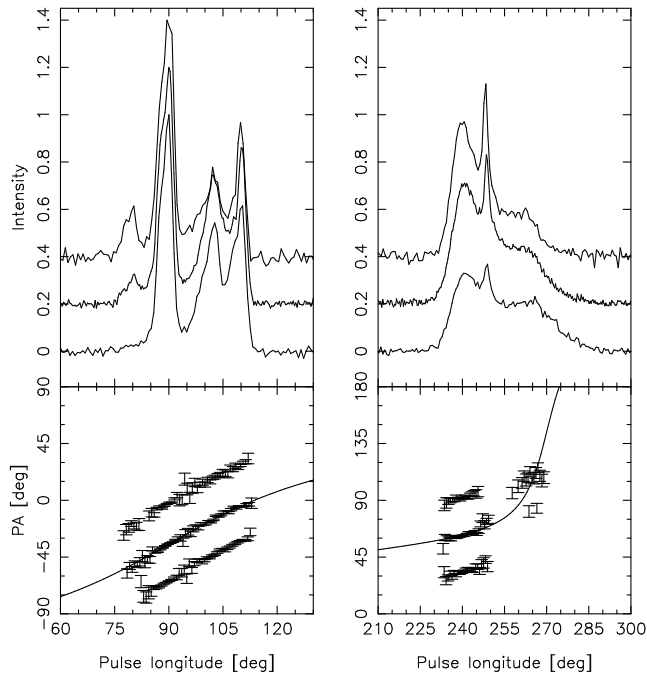

\begin{center}
\includegraphics[width=1.05\hsize,angle=270]{1055profilesA.ps}\hspace*{0.05\hsize}\includegraphics[width=1.05\hsize,angle=270,trim=0 20 0 0, clip=true]{1055profilesB.ps}
\end{center}
\caption{\label{Fig1055profiles}{\em Top panels:} The pulse profiles
of PSR~B1055--52 at three observing frequencies drawn with different
vertical offsets. From top to bottom the wavelength are 10~cm, 20~cm
and 50~cm. {\em Bottom panels:} The PA of the linear polarization and
an RVM fit. The offsets are the same as in the top panels. These are
the August 2006 observations. }
\end{figure}

The pulse profiles of PSR~B1055--52 at different frequencies are shown
in Fig. \ref{Fig1055profiles}. The profiles are aligned and normalized
using the peak of component II which is put at \degrees{90} pulse
longitude. One can see that there is not much evidence for frequency
evolution in the relative position of the components, but the peak
amplitudes do evolve with frequency. In the MP the first component
disappears at 50~cm while components III and IV become relatively
stronger. The IP shows a similar frequency evolution with the leading
half of the profile becoming weaker relative to the trailing
component. In relative terms, the IP itself is strongest at high
frequencies. While the components of the IP are sharper at 10~cm this
is less clear in the MP. Notice also that component VII is the only
component which changes its centroid position with respect to the
other components.

There is also no evidence of the full widths of the MP and IP (or
their separation) being frequency-dependent, potentially an important
hint as to the pulsar's geometry. A profile at 170~MHz, ascribed to
McCulloch et al. and published in \cite{ran83}, shows a merging of
components III and IV in the MP, but little overall widening of the
profiles.

\section{Geometric implications}
\label{SectGeometry}

The PA-swing is plotted in the bottom panel of
Fig. \ref{Fig1055profiles} for the three frequencies. We found a
significant offset of the 50~cm data with respect to the other
wavelengths when we use the rotation measure (RM) of $47.2\pm0.8$
rad m$^{-2}$ \citep{tml93}. By applying an RM of 46 rad m$^{-2}$ the
PA-swings at the different frequencies line up, suggesting that the
shape of the PA-swing, like that of the profile, is independent of
frequency. This value of the RM is consistent with the value of
$44\pm2$ rad m$^{-2}$ measured by \cite{njk+08}.

The shape of the PA-swing can be used to constrain the angles between
the magnetic and rotation axis ($\alpha$) and the angle between the
line of sight and the rotation axis ($\zeta=\alpha+\beta$, where
$\beta$ is the impact parameter).  According to the rotating vector
model (RVM; \citealt{rc69a}) the PA ($\psi$) can be described as a
function of the pulse longitude $\left(\phi\right)$ by
\begin{equation}
\label{EqRVM}
\tan\left(\psi-\psi_0\right)=\frac{\sin\alpha\;\sin\left(\phi-\phi_0\right)}{\sin\zeta\;\cos\alpha-\cos\zeta\;\sin\alpha\;\cos\left(\phi-\phi_0\right)},
\end{equation}
where $\psi_0$ and $\phi_0$ are the PA and the pulse longitude
corresponding to the inflection point of the PA-swing. Using this
definition implies that we follow the observer's sign convention for
the PA, which differs from that introduced by \cite{dt92} and
\cite{ew01}.

If the emission height $r$ is zero, or at least negligible compared
with the light cylinder radius $R_{LC}$, the inflection point of the
PA-swing coincides with the fiducial point (defined as the pulse
longitude where the magnetic pole at the surface is directed towards
us).  However, for non-zero emission heights this is no longer the
case.  As the emission height increases the total intensity appears to
shift towards earlier pulse longitudes due to aberration (by
$r/R_{LC}$) and the decreasing light travel distance to the observer
(also by $r/R_{LC}$), while the PA-swing shifts to later pulse
longitudes with respect to the fiducial point (by $2r/R_{LC}$). Thus
the total relative shift between the steepest gradient of the PA-swing
with respect to the emission beam's centroid is given by
\begin{equation}
\label{EqBCW}
\Delta \phi_\mathrm{BCW} = 4r/R_{LC}
\end{equation}
(\citealt{bcw91,dyk08}), where $r$ is assumed to be small compared
with $R_{LC}$.

In PSR~B1055--52 the degree of linear polarization is low in the
trailing half of the IP, but fortunately it is enough to obtain the
PA. This means we have a full PA coverage at virtually every pulse
longitude where emission is detected at a far better time resolution
than was available to \cite{lm88}. This enables us to re-examine and
place constraints on the geometric solution of this pulsar.  In this
section we formally confirm that a nearly-orthogonal solution is
preferable to a nearly-aligned solution and exploit the $\gamma$-ray
profile to establish an integrated geometry.

\subsection{Nearly-aligned rotator?}

We first consider the possibility that PSR~B1055--52 might be a
nearly-aligned rotator. In such a geometry the MP and IP originate
from the two edges of a hollow cone-like beam from a single pole
(e.g. \citealt{ml77}). The separation between the two components can
be large, but, as observed, will not necessarily be \degrees{180} in
pulse longitude. The large widths of the components together with the
steep trailing edge of the MP and the leading edge of IP are also
typical for such geometry, making this model a promising possibility
to consider.

However, for a nearly-aligned rotator one expects the PA-swing to be
flatter than is observed. In the RVM model the steepest gradient is
associated with the line of sight passing the fiducial plane. In the
case of an aligned rotator this plane should be at a pulse longitude
between the MP and IP, hence the observed PA-swing will be relatively
flat. As has been pointed out (by \citealt{wqxl06} for example), the
steepness of the PA-swing observed for both the MP and IP rules out
the one pole geometry. It should be added that this conclusion is also
valid even when \degrees{90} jumps in the PA-swing caused by different
plasma modes dominating in different parts of the pulse profile
(e.g. \citealt{mth75}) are considered.

The PA-swing inflection point can in principle be shifted from its
unshifted pulse longitude between the MP and IP (as would be expected
for an aligned rotator) to appear at the pulse longitude of one of the
pulse components (MP or IP) when the aberrational shift of
Eq. \ref{EqBCW} is considered. This effect could therefore in
principle make the PA-swing steep in one of the two
components. However, it cannot make them steep in both since the RVM
model predicts a steep gradient at only one point in the PA-curve,
thereby failing to explain the observed PA-swing.

\subsection{Nearly orthogonal rotator}
\label{SctOrthogonal}

\begin{figure}
\includegraphics[height=0.99\hsize,angle=270,trim = 13mm 0mm 0mm 0mm, clip]{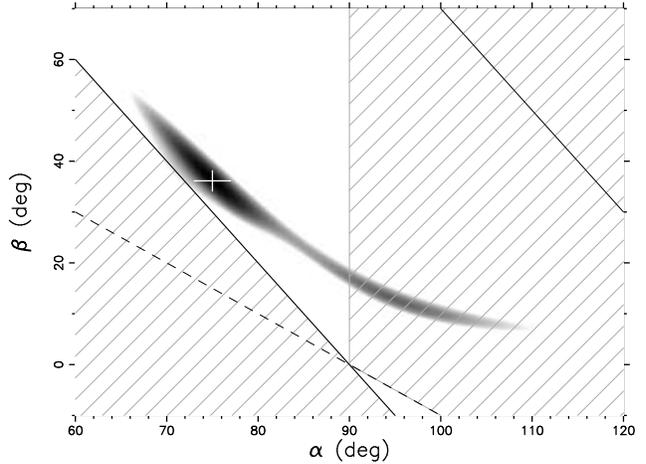}
\caption{\label{Fig1055chi2}The reduced $\chi^2$ grid obtained by
fitting the RVM to the PA-swing of both the MP and IP of the 1369~MHz
timing data. Black corresponds to 2.7 and white to 20. The cross
indicates the position of the minimum of the $\chi^2$ surface
($\alpha_\mathrm{MP}=\degrees{75}$ and
$\beta_\mathrm{MP}=\degrees{36.1}$). Solutions to the right of the
right hand side diagonal solid line are excluded because they require
a $\beta_\mathrm{IP} > \degrees{90}$ and solutions to the left of the
left hand side diagonal solid line are excluded because they require a
negative $\beta_\mathrm{IP}$ (corresponding to a declining PA-swing
for the IP). Solutions below the dashed line (for
$\alpha_\mathrm{MP}<\degrees{90}$) and above the dashed line (for
$\alpha_\mathrm{MP}>\degrees{90}$) are excluded as they do not allow
the $\gamma$-ray emission to be emitted from above the null line. }
\end{figure}

In searching for a self-consistent solution, we follow \cite{wqxl06}
and consider a model where the profile of PSR~B1055--52 is generated
by emission from an orthogonal rotator.  

Although \cite{mhma78} claimed that the PA-swing cannot be
extrapolated between the MP and IP, we (like \citealt{wqxl06} with
inferred data from \citealt{lm88}) had no problem in fitting the
PA-curve with the RVM model (Eq. \ref{EqRVM}). In fact, the reduced
$\chi^2$ of 2.7 is quite good, although not perfect
(Fig. \ref{Fig1055chi2}).  The lowest $\chi^2$ is found for
$\alpha_\mathrm{MP}=\degrees{75}$ and
$\beta_\mathrm{MP}=\degrees{36.1}$ (where $\alpha_\mathrm{MP}$ and
$\beta_\mathrm{MP}$ are the MP values of $\alpha$ and $\beta$),
suggesting that the orientation of the magnetic axis is indeed close
to orthogonal and in agreement with \cite{wqxl06}. This implies
$\alpha_\mathrm{IP} = \degrees{180}-\alpha_\mathrm{MP}=\degrees{105}$
and an impact parameter at the IP of
$\beta_\mathrm{IP}=\beta_\mathrm{MP}+2\alpha_\mathrm{MP}-\degrees{180}=\degrees{6.1}$.
Note that these fits require that all the emission at each pole comes
from the same height, a point to which we will return in
Sect. \ref{SectEmissionheight}. The best fit is shown in
Fig. \ref{Fig1055profiles} as the solid curve. The biggest discrepancy
between the fit and the data occurs in the trailing component of the
IP, where the fit is too flat compared with the model.  The steepest
gradient in the MP ($\phi_0$ of Eq. \ref{EqRVM}) is found exactly at
the location of the highest peak in the pulse profile
($\phi=\degrees{90}$ in the plots).

The fact that the PA-swing can be extrapolated between the MP and IP
further implies that the emission heights at both poles cannot be too
large. If there were a large differential emission height, the
PA-swings of the MP and IP would be shifted by different amounts
(\citealt{bcw91,ha01,dyk08}), making it impossible to fit the PA-swing
with a single RVM curve. By modelling these relative shifts we find
that the lowest $\chi^2$ is found when the emission height of the IP
is higher than that of the MP by $0.07R_\mathrm{LC}$ ($\approx{700}$
km) with an uncertainty of $0.10R_\mathrm{LC}$ ($\approx{1000}$ km),
which is consistent with the emission height being identical at both
poles.  The uncertainty is derived by finding the differential
emission height at which the reduced $\chi^2$ of the PA fit has
doubled. Although the solution found by \cite{wqxl06} (who find an
IP/MP emission height differential of between $0.10R_\mathrm{LC}$ and
$0.20 R_\mathrm{LC}$) is in the allowed range, we will argue later
that this solution is very unlikely. We therefore seek a solution
with identical emission heights for the MP and IP.

\subsection{Implications of the $\gamma$-ray profile}

As \cite{lm88} pointed out, and as is evident in
Fig. \ref{Fig1055chi2}, there are RVM solutions over a wide range of
$\alpha_\mathrm{MP}$ values which have a comparable $\chi^2$. However,
we can utilize additional constraints to confirm that we have the
right solution. By assuming that $\beta_\mathrm{IP}$ must be smaller
than \degrees{90} (or else the full opening angle of the radio beam
would be larger than \degrees{180}) and that $\beta_\mathrm{IP}$ must
be positive (or else the PA-swing of the IP would decrease) only
solutions between the two diagonal solid lines in
Fig. \ref{Fig1055chi2} are allowed.

A further and critical constraint can come from the observed phase
and width of the $\gamma$-ray profile (\citealt{tbb+99}). Different
magnetospheric locations are proposed for the $\gamma$-ray emission of
pulsars. Early models put the acceleration zones at low altitudes
above the magnetic poles (polar cap models; e.g. \citealt{dh96}). More
recently, it has been argued that the emission height could be
considerably larger for fieldlines close to the last open fieldlines
giving rise to the ``slot gap'' model (e.g. \citealt{mh04a}) and the
``two-pole caustic'' model \citep{dr03}. A separate class of models
are the so-called ``outergap'' models \citep{chr86b,rom96a}, which
place the acceleration zones above the ``null lines''
\citep{hol73}. These are the lines on which the charge density of a
pulsar magnetosphere changes sign (\citealt{gj69}).

Numerical simulations show that only an outer gap model can reproduce
the observed shape of the $\gamma$-profile of PSR~B1055--52 to a good
approximation \citep{wrwj08}. The two-pole caustic model, which has
$\gamma$-ray production below the null line, predicts in general
profiles with observed emission from both magnetic poles, which makes
it difficult to reproduce the observed $\gamma$-ray pulse width. In
order to reproduce the observed single-peaked $\gamma$-ray profile the
two-pole caustic model predicts either $\alpha$ or $\zeta$ values
smaller than allowed by the PA-swing parameters between the two solid
diagonal lines of Fig. \ref{Fig1055chi2}. Polar cap models, which have
low emission heights, also require too small $\alpha$ and $\zeta$
values in order to get a broad enough $\gamma$-ray profile. On the
other hand outer gap models predict, by definition, that only
emission from one pole can be seen for any given observer, thereby
qualitatively reproducing the observed $\gamma$-ray profile for a wide
range of geometries.

In order to reproduce the observed width of the $\gamma$-ray profile
the gap thickness $w$ (the fraction of the angle from the last closed
fieldline to the magnetic axis) should be of the order of $0.18$
(assuming that the $\gamma$-ray efficiency is equal to $w$), showing
that the gap is very large in comparison with most $\gamma$-ray
pulsars \citep{wrwj08}. Large outer gaps are typical for highly
efficient old $\gamma$-ray pulsars (e.g. \citealt{rom96a}). The outer
gap model further predicts that two caustics are formed at the edges
of the $\gamma$-ray beam, which could account for the sub-structure
visible in the EGRET data.

\begin{figure}
\includegraphics[height=0.99\hsize,angle=270,trim = 13mm 0mm 0mm 0mm, clip]{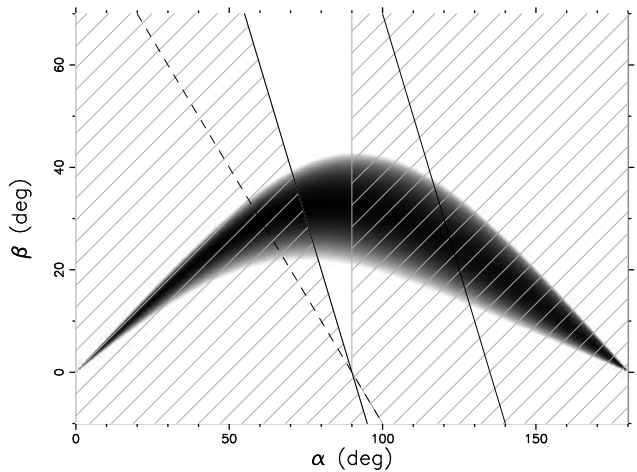}
\caption{\label{Fig1055chi2_2}The reduced $\chi^2$ grid obtained by
fitting only the PA of the 1369~MHz timing data of the MP
region. Black corresponds to 1.1 and white to 3.0. For an explanation
of the lines see Fig. \ref{Fig1055chi2}.}
\end{figure}

If we accept that the $\gamma$-rays are produced above the null
surface of the MP (i.e. they are produced by the outer gap) large
parts of the $\alpha_\mathrm{MP}-\beta_\mathrm{MP}$ plane can be ruled
out.  Together with the $\degrees{0}<\beta_{IP}<\degrees{90}$
restriction we can eliminate the entire hashed regions in
Fig. \ref{Fig1055chi2} from consideration. As one can see in
Fig. \ref{Fig1055chi2_2}, using these constrains the geometry of
PSR~B1055--52 is already constrained even without fitting the shape of
the PA-swing of the IP as there is only a relatively small wedge
shaped region of allowed solutions in the plot. This is important,
because it shows that our conclusions do not depend strongly on the
shape of the PA-swing of the IP which deviates slightly from the RVM
prediction. We will therefore adopt the RVM solution with the lowest
$\chi^2$ in this paper which has $\alpha_\mathrm{MP}=\degrees{75}$ and
$\beta_\mathrm{MP}=\degrees{36.1}$.

\subsection{The geometric picture}

\begin{figure}
\begin{center}
\includegraphics[width=0.99\hsize,angle=0]{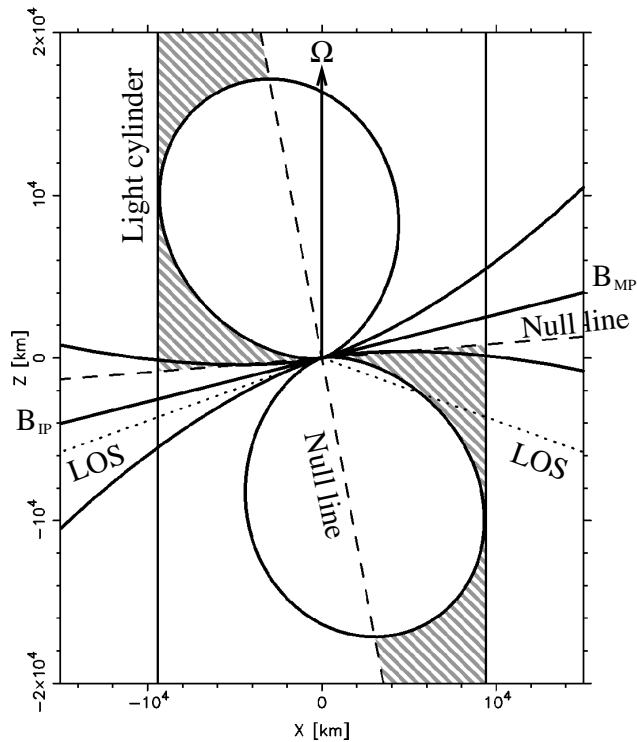}
\end{center}
\caption{\label{Fig1055field}The geometry of the dipole field of
PSR~B1055--52 assuming $\alpha_\mathrm{MP}=\degrees{75}$. The rotation
axis points upwards and the light cylinder is indicated by the
vertical lines. The line of sight (dotted line ``LOS'') makes a large
angle (\degrees{36.1}) with the magnetic axis of the MP (right hand
side of the plot), while this angle is much smaller (\degrees{6.1})
for the IP (left hand side of the plot). The dashed lines indicate the
location of the null surface, and the last closed fieldlines which
conventionally define a corotating region are shown. At the MP we must
be viewing fieldlines which lie above the null surface (in the shaded
region) -- and which will be well within the corotating region if the
radio emission is, as usually assumed, close to the surface. The
$\gamma$-ray profile, which is observed to overlap the MP radio
emission, is most likely formed in the shaded area, which
indicates the volume above the null surface where an outer gap could
potentially be formed.}
\end{figure}

The emission geometry in the fiducial plane is illustrated in
Fig. \ref{Fig1055field}.  One can see that the line of sight passes
the magnetic axis of the IP much more closely than to the pole of the
MP, consistent with the steeper slope of the PA-swing of the IP. Thus,
the IP emission must arise from open fieldlines as is generally
thought to be the case for radio pulsars. However, the incidence of
the line of sight with the magnetic axis of the MP is very large,
which immediately leads to one of two possibilities. If we require the
emission to arise from open fieldlines, then it must originate at a
considerable fraction of the light cylinder. Otherwise we are forced
to accept emission from presumed closed fieldlines at low altitudes.

Fig. \ref{Fig1055field} also shows the null lines (dashed). These are
the $\mathbf{\Omega}\cdot\mathbf{B}=0$ lines on which the charge
density of the magnetosphere changes sign. The tangent of the
fieldlines are perpendicular to the rotation axis at the point where
they intersect the null line. It immediately follows that for the
assumed geometry we `see' fieldlines above the null line for the MP
(Fig. \ref{Fig1055field}), which suggests that the radio MP is
produced in a region where an outergap is likely to exist. It can be
shown that this is a general rule. In any pulsar where emission from
both poles are detected, at one of the poles our line of sight must
see emission from the region above the null line, i.e. the likely site
of any ``outergap''. Assuming that this outergap is the site of the
observed $\gamma$-ray emission, it is no surprise that the phase range
of the $\gamma$-ray profile overlaps that of the MP rather than
the IP. We note that were we to choose an RVM solution with
$\alpha_\mathrm{MP}>\degrees{90}$, then the IP would be the pole which
views the outergap region, and the $\gamma$-ray profile would be at
the wrong phase.

Most other known $\gamma$-ray pulsars are, unlike PSR~B1055--52,
typically found to have a single peaked radio profile with a large
offset from the $\gamma$-ray profile. The lines of sight of these
pulsars are also believed to have a large $\beta$ value for one of the
poles. This causes the line of sight to miss the radio beam of the
magnetic pole for which the $\gamma$-rays are observed, hence a large
offset is seen between the radio and the $\gamma$-rays. PSR~B1055--52
is different because, for some reason, it has a large enough radio
beam to be seen by a line of sight with a large angle $\beta$.

One further conclusion can also be drawn. With $\beta_\mathrm{MP}$
being so large at the MP ($\degrees{36.1}$), the corresponding
$\beta_\mathrm{IP}$ at the IP will be {\em positive} (since
$\beta_\mathrm{IP}=\beta_\mathrm{MP}+2\alpha_\mathrm{MP}-\degrees{180}=\degrees{6.1}$). This
implies that at the IP we view the pole on the side {\em away} from
the null line, and therefore do not see fieldlines which thread the
outergap region.

\section{Beam illumination}
\label{SectBeamIllumination}
\subsection{Beam mapping}

We now use the widths of the MP and IP pulse profile to place
additional constraints on the magnetosphere's geometry. The first
point that needs to be considered is the deviation of the MP-IP
separation from \degrees{180} in pulse longitude. This could in
principle be caused by a differential emission height, but, as we
argue in Sect. \ref{SectEmissionheight}, this solution is very
unlikely. This leaves us with the less satisfactory conclusion that
at least one of the radio beams is not fully active
(e.g. \citealt{lm88,kj07}).

Our next aim will be to investigate which parts of the MP and IP beam
should be active in order to explain the observations. This is done by
mapping the observed pulse longitudes onto the fieldlines which are
pointing in our direction.  The pulse longitudes of the edges of the
MP and IP are therefore an important factor in determining which
fieldlines are active. We take the positions of the steepest gradient
of the RVM fit ($\phi_0$ and $\degrees{180}+\phi_0$, see
Eq. \ref{EqRVM}) as the reference. These are located in component II
of the MP and at the trailing edge of the IP. The emission of the MP
is located at a pulse longitude range of
$\Delta\phi_\mathrm{MP}=\degrees{-15}$ to \degrees{+22} with respect
to the first reference point and the emission of the IP confined to
$\Delta\phi_\mathrm{IP}=\degrees{-42}$ to \degrees{+10} using the
second.

This can be used to identify the fieldlines which are active after
correcting for the relative shift between the PA-swing (which is used
as reference) and the total intensity for a non-zero emission height
(Eq. \ref{EqBCW}). This means the true longitude of the emission point
with respect to the fiducial point in the corotating frame is
$\Delta\phi+\Delta\phi_\mathrm{BCW}$. If $\rho$ is the opening angle of the
fieldline at the emission height and $\varphi$ is the azimuthal angle
of the fieldline with respect to the magnetic axis such that zero
corresponds to the fiducial plane, these are given by
\begin{eqnarray}
\label{EqTranform1}
\cos\rho = \cos\alpha \cos\zeta+\sin\alpha \sin\zeta \cos\left(\Delta\phi+\Delta\phi_\mathrm{BCW}\right),\\
\label{EqTranform2}
\sin\varphi=\frac{\sin\zeta\sin\left(\Delta\phi+\Delta\phi_\mathrm{BCW}\right)}{\sin\rho}
\end{eqnarray}
where $\Delta\phi=\phi-\phi_0$ is the observed pulse longitude
difference with respect to the reference point as discussed above
(\citealt{ggr84,ggk+04,dyk08} see also Appendix \ref{SctApp} for a
derivation). 

Note that $\rho$ is only identical for the leading and trailing edge
of a beam when the beam is symmetric about the fiducial plane. In
that case
$\Delta\phi_\mathrm{leading}+\Delta\phi_\mathrm{BCW}=-\left(\Delta\phi_\mathrm{trailing}+\Delta\phi_\mathrm{BCW}\right)=\frac{1}{2}W$,
where $W$ is the full pulse width.

Assuming a dipolar field, the opening angle of the fieldline $\rho$
can be expressed in terms of the polar angle $\theta$ (colatitude) of
the point of emission \citep{gg01}
\begin{equation}
\tan\theta = -\frac{3}{2\tan\rho}+\sqrt{2+\left(\frac{3}{2\tan\rho}\right)^2},
\end{equation}
which then can be expressed in terms of the footprint parameter $s$ on
the polar cap
\begin{equation}
s = \sin\theta\sqrt{\frac{R_\mathrm{LC}}{r}}.
\end{equation}
The magnetic axis corresponds to $s=0$, and $s=1$ corresponds
to last open fieldlines. As will be discussed in
Sect. \ref{SectPolarMap}, the polar cap is not necessarily strictly
circular if the magnetic axis makes an angle with the rotation axis.

\begin{figure*}
\includegraphics[height=0.45\hsize,angle=270]{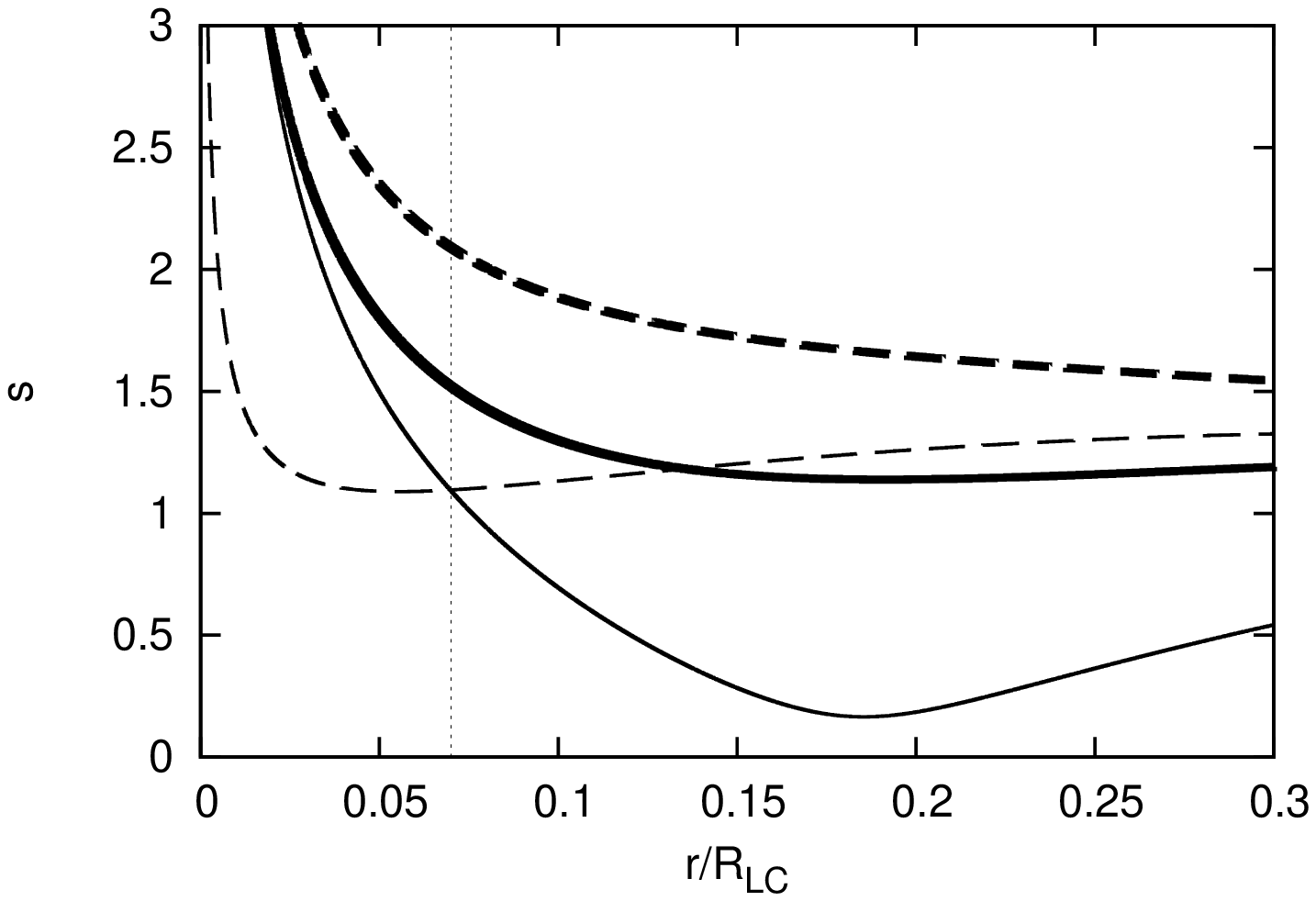}\hspace*{0.05\hsize}\includegraphics[height=0.45\hsize,angle=270]{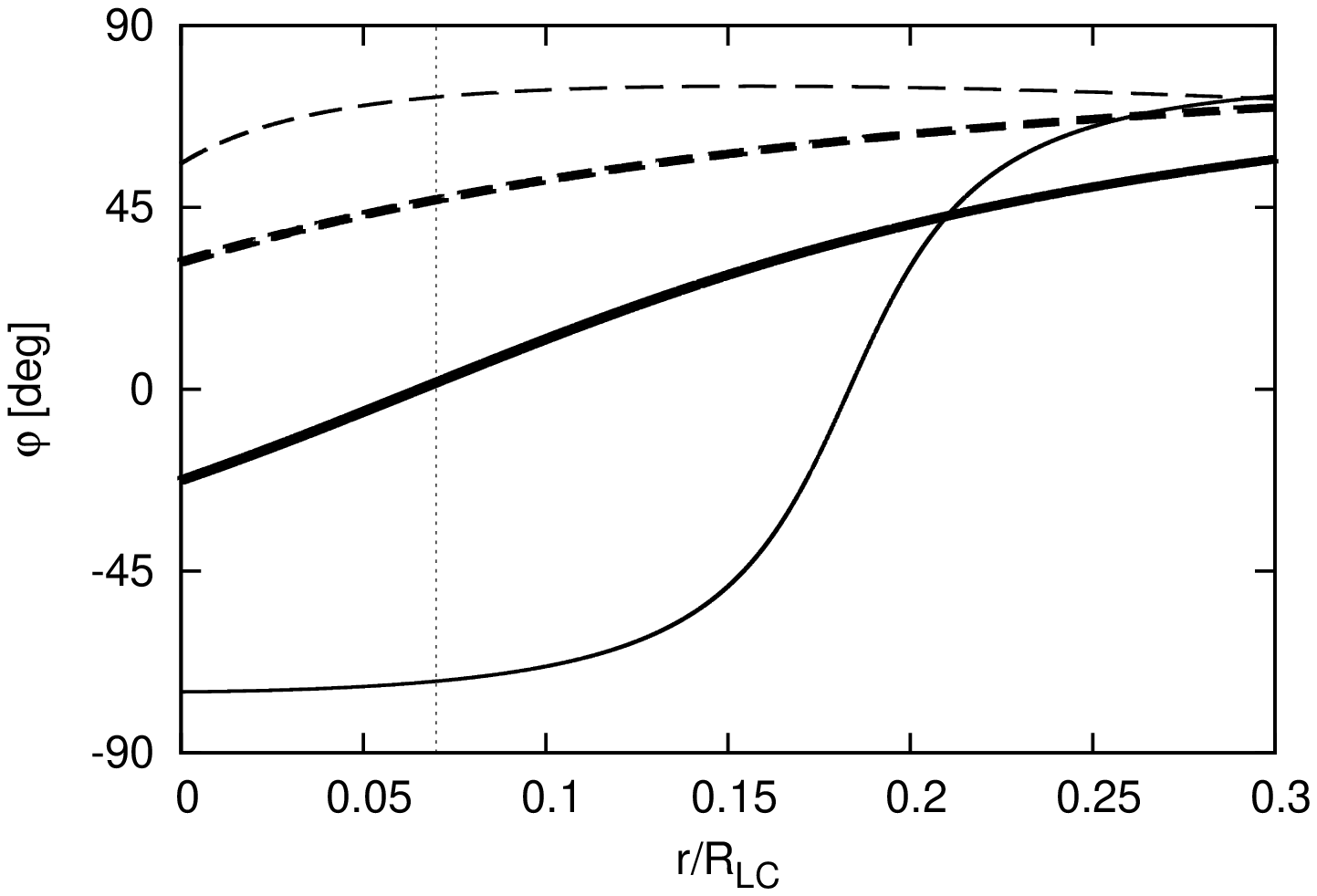}
\caption{\label{Fig1055illumination}The geometrical solution for which
fieldlines should be active in order to fit the PA-swing and the pulse
profiles as function of the emission height. The fieldline is
characterized by the footprint parameter $s$ and the azimuthal angle
$\varphi$. The solution for the MP are the thick curves and that of
the IP are the thinner curves (assuming
$\alpha_\mathrm{MP}=\degrees{75}$). The solid lines indicate the
solution for the leading edge of the component and the dashed line
that of the trailing edge. The dotted line indicates
$r=0.07R_\mathrm{LC}$. }
\end{figure*}

The geometrical solution for the leading and trailing edges of the MP
and IP are shown in Fig. \ref{Fig1055illumination} as a function of
the emission height (because $\Delta\phi_\mathrm{BCW}$ is a function
of the emission height). These plots identify which part of the polar
cap should be active in order to explain the component widths and
separations in combination with the shape and shift of the
PA-swing. One can see that $s$ becomes very large when the emission
height is very low. This is because there is a trade-off between the
emission height and the $s$ value in order to get the required
observed component width. At large emission heights there is a
turnover and $s$ increases with emission height. This can be
understood in terms of the shift of the PA-swing with respect to the
emission profile.
In order to prevent the PA-swing from having its steepest gradient far
outside the beam, $s$ has to increase. At the same time the fraction
of the beam which is active has to decrease in order to avoid
generating profile components which are larger than observed.  This
can be seen in the $\varphi$ curves, which approach each other for
large emission heights such that only the trailing edges of the beams
are active.

\subsection{The emission height}
\label{SectEmissionheight}

It is not possible to determine the emission height based just on the
shape of the pulse profile in combination with the shape of the
PA-swing. Additional assumptions are required to constrain the
emission height further. \cite{wqxl06} make two assumptions in order
to derive their solution. Firstly, both the MP and IP beam are assumed
to be filled (i.e. that $\varphi$ of the leading and trailing edge
have the same magnitude but opposite sign). This choice is not
available to us because we have insisted on a solution fully
consistent with retardation and aberrational effects. The second
assumption made by \cite{wqxl06} is that the $s$
parameter\footnote{The $s$ parameter corresponds to $\eta$ in
\cite{wqxl06}.} of the MP is assumed to be 1. One can see in
Fig. \ref{Fig1055illumination} that the trailing edge (dashed line) of
the IP and especially the MP have to be emitted from fieldlines
further away than $s=1$ from the magnetic axis (which is true for all
possible emission heights). This shows again that the solution
presented by \cite{wqxl06} is not self-consistent.

Fig. \ref{Fig1055illumination} shows that the solid and dashed $s$
curves of the IP cross at $r=0.07R_\mathrm{LC}$. This corresponds to a
beam which is completely filled such that the distances of the
footprints of the active fieldlines to the magnetic axis are the same
for the leading and trailing edge. This is consistent with the
$\varphi$ values at this height which have the same magnitude but an
opposite sign. This happens at an $s$ value slightly larger than 1,
indicating that the beam is slightly bigger than the conventional
polar cap size. The same emission height corresponds to a special
geometry for the MP as the solid $\varphi$ line crosses zero at that
height. This means that the MP beam is exactly half filled for
the emission height at which the IP is exactly completely
filled. This arises because at this emission height
$\phi_\mathrm{leading}+\Delta\phi_\mathrm{BCW}=-\left(\phi_\mathrm{trailing}+\Delta\phi_\mathrm{BCW}\right)$
for the IP.  For the same emission height at the MP
$\phi_\mathrm{leading}+\Delta\phi_\mathrm{BCW}=0$. There is no obvious
reason why an exactly half filled MP should have physical
significance, so it is presumably a coincidence this solution occurs
at the same emission height as the fully filled IP solution. Moreover,
we cannot know if the emission height is exactly
$r=0.07R_\mathrm{LC}$.

The trailing edge of the radio MP and the leading edge of the radio IP
are relatively steep (Fig. \ref{Fig1055profiles20}). This suggests
that we only see the trailing side of the radio MP and only the
leading side of the IP (e.g. \citealt{lm88}).  If we accept this idea,
then the emission height must be smaller than $0.07R_\mathrm{LC}=700$
km, because for higher emission heights the IP's trailing edge (rather
than leading) would correspond to an edge of the beam (see
Fig. \ref{Fig1055illumination}). This would imply that the radio
emission height of PSR~B1055--52 is not abnormal compared with other
pulsars (e.g. \citealt{bcw91,mr01,wj08b,kg97}).  

If there is a differential emission height between the two poles, then
that of the IP is larger (see Sect. \ref{SctOrthogonal}). However, the
IP's emission height cannot be larger than $0.07R_\mathrm{LC}$, or
else its leading edge would no longer be the edge of the beam. The
emission height of the MP could be slightly lower, although not too
much as it would make its $s$ value unrealistically large. Therefore
roughly similar emission heights at $0.07R_\mathrm{LC}$ (or slightly
lower) is the most likely solution.

\subsection{Polar map}
\label{SectPolarMap}

\begin{figure}
\includegraphics[height=0.99\hsize,angle=270]{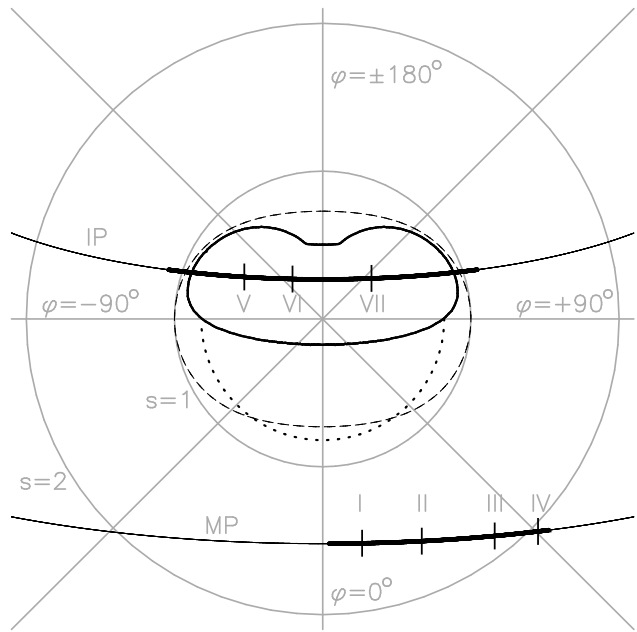}
\caption{\label{FigPolarMap}The polar cap map of PSR~B1055--52 based
on the assumption that both poles have identical structure so that we
can ``reflect'' the observed IP mapping in the horizontal axis. The
conventional polar cap size of an aligned rotator is indicated by the
circle with radius $s=1$. The dashed ellipse shows the polar cap
including meridional compression.  The fieldlines inside the
kidney-shaped area are open and do not pass through any null line. The
top and bottom solid curves are the lines of sight for the MP and IP
respectively, where the active portions (assuming an emission height
of $0.07R_\mathrm{LC}$) are the thicker parts of the curves. The
location of the peaks are indicated as well (see
Fig. \ref{Fig1055profiles20}). The dotted semi-circle indicates
approximately the footprints of the fieldlines on which $\gamma$-rays
are thought to be produced. }
\end{figure}

Having chosen a preferred (maximum) height, we may use the
geometric solution of Fig. \ref{Fig1055illumination} to generate a map
of the polar cap of PSR~B1055--52. The result is shown in
Fig. \ref{FigPolarMap} and combines the IP and MP on a single pole,
thereby assuming that the emission structure at both poles is
identical. Obviously, because we only have information about the line
of sight which passes over different parts of the polar cap of the MP
and IP, we cannot know whether this assumption is true. It must be
borne in mind that both the IP and MP are observed on the same
meridional planes of $\varphi$ (i.e. they both are viewed below the
magnetic axis in Fig. \ref{Fig1055field}), and in
Fig. \ref{FigPolarMap} the IP line of sight traverse has been
``reflected'' in the horizontal axis.  This reflection ensures that a
point in the polar map is associated with fieldlines which have
identical geometric properties for both the MP and IP. The circle
$s=1$ corresponds to the footprints of the last closed fieldlines of
an aligned dipole touching the light cylinder. For non-aligned dipoles
the polar cap shape gets distorted by meridional compression
(e.g. \citealt{big90b}). The predicted shape for a dipole field
inclined at $\degrees{75}$ is shown as the dashed oval in
Fig. \ref{FigPolarMap}. However, it must be noted that even models which
take into account the distortion of the magnetic field lines close to
the light cylinder predict roughly circular beams
(e.g. \citealt{ry95}).

In Fig. \ref{FigPolarMap} we also indicate the kidney-shaped region
within which fieldlines of a dipole inclined at $\degrees{75}$ do not
cross the null surface inside the light cylinder.  This shape is
numerically derived by determining for each dipole fieldline whether
the null surface $\mathbf{\Omega}\cdot\mathbf{B}=0$ is penetrated
within the light cylinder. It can also be determined analytically
(\citealt{qlw+04}). In the upper part of the polar map the null
surface will lie at a great distance from the polar region (see
Fig. \ref{Fig1055field}), but in the lower section it may be close
enough to the star to set up an active outergap. We also indicate the
approximation of the footprints of the $w=0.18$ fieldlines, argued by
\cite{wrwj08} to be the fieldlines on which the $\gamma$-ray emission
is produced (dotted semi-circle in Fig. \ref{FigPolarMap}).

Note that the locus of active fieldlines on which the IP is generated
is symmetric about $\varphi=0$ and barely extends outside
$s=1$. Furthermore, the IP fieldlines, apart from their outer edges,
are contained within the kidney-shaped locus of the open fieldlines
which do not cross a null surface. It was suggested by \cite{gj69}
that such fieldlines carry the outflowing current from the pulsar.

By contrast, the footprints of the active MP fieldlines lie well
outside $s=1$, extending to radii more that double this
value. Strangely, they cover only the latter half of a
symmetrically-defined full beam. However, as pointed out above, the
MP's fieldlines will thread the null surface close to the star's
surface (at about 160 km) and the emission will occur in a plasma
region usually considered to be inactive and corotating, so we can be
sure that the MP is not a traditional polar cap profile.

The solution shown is for $r=0.07R_\mathrm{LC}$, which is argued
to be the maximum allowed emission height
(Sect. \ref{SectEmissionheight}). For smaller emission heights both
the line of sight of the MP and the IP will move to larger values of $s$
(see Fig. \ref{Fig1055illumination}) and the active parts of the lines
of sight will start leftwards of their current location (more negative
values of $\varphi$). 

\section{Discussion}
\label{SectDiscussion}

In our study of PSR~B1055--52 we have focused on investigating the
geometry of its magnetosphere. Its wide MP and IP, together with a
clearly-positioned single gamma-ray peak, place considerable
constraints on any potential model. In an earlier attempt to locate
the emission regions, \cite{wqxl06} fitted an RVM model to published
figures (\citealt{lm88}) for the pulsar's PA-swing and did not
consider the constraints that aberration and retardation might place
on the choice of emission heights. We have been able to exploit new
data, gathered at Parkes in recent years, to verify the RVM fit and to
attempt a self-consistent model.

One of our principal conclusions is that both the MP and IP are
emitted at roughly the same height. The optimum fit
($\alpha_\mathrm{MP}=\degrees{75}$,
$\beta_\mathrm{MP}=\degrees{36.1}$) is in line with earlier results.

This PA fit has a number of interesting implications for PSR~B1055--52
which are independent of assumptions about the emission heights. Since
$\zeta_\mathrm{MP}$ (i.e. $\alpha_\mathrm{MP}+\beta_\mathrm{MP})$, at
$\degrees{111}$ is considerably greater than $\degrees{90}$, our
line of sight at the MP must see emission on fieldlines at locations
which are above the ``null'' line (Fig. \ref{Fig1055field}), in or
beyond territory normally ascribed to an outergap (as was also noted
by \citealt{hs02b}). In fact, on elementary geometric grounds, it must
be true for every pulsar with a detected IP that at one pole we are
viewing above the null line and at the other below it --- a point
not widely realised before.

However, in the case of PSR~B1055--52 (but not necessarily for all
pulsars with IP's), $\zeta_\mathrm{IP}>\alpha_\mathrm{IP}$ and so our
line of sight at the IP must cross the magnetic axis on fieldlines
which do not thread the outergap at the opposite pole
(Fig. \ref{FigPolarMap}). On these grounds alone, we can expect that
emission at the MP and IP will reflect the different conditions of
these two regions. This can be contrasted with the emission from
PSR~B1702--19, where at the MP, in one of its two possible solutions,
$\alpha_\mathrm{MP}=\degrees{99}$ and
$\zeta_\mathrm{MP}=\degrees{91.5}$ \citep{wws07}. Then, at one pole
(IP) we must be viewing fieldlines above the null line and at the
other (MP) equivalent fieldlines before they cross the null line, with
a greater likelihood of coordinated emission.

Although the radio emission at both poles must come from about the
same height, the precise height is not easy to determine. To
first-order, aberration at a fixed height simply shifts the entire PA
curve to later pulse longitudes by an amount proportional to the
height. Thus, as long as the emission height is small compared to the
light cylinder radius and unchanged across the profile, we have no way
of knowing from the shape of the PA curve by how much it has been
shifted. The exceptionally wide MP and IP lead us to believe that,
whatever the assumed height, it is impossible to confine the
footprints of the emitting regions to a polar cap of conventional
dimensions (i.e. defined by last closed dipole fieldlines
touching the light cylinder). This requirement is particularly acute
for the MP, whose trailing components must lie on fieldlines located
about twice the conventional polar radius from the pole. Furthermore,
since the fiducial point of the MP profile must lie ahead of the
steepest gradient of the PA-swing, the observed MP width will never
occupy much more than half its notional beam.  Choosing a height less
than about 500 km would force the polar cap size to escalate rapidly
(see Fig. \ref{Fig1055illumination}) and therefore a balance has to be
struck.  Here a height 7 per-cent of the light cylinder radius is
assumed, so that the IP emission fills its beam and has footprints at
around $s=1$ on the opposite side of the pole (assuming identical
polar caps at both poles). This choice also has the advantage that the
steep edges of the MP (trailing) and IP (leading) correspond to beam
edges (see Figs. \ref{Fig1055illumination} and \ref{FigPolarMap}). We
therefore conclude that the emission height is unlikely to be higher
or much lower than 700 km, a typical value for pulsars.

\begin{figure}
\includegraphics[height=8cm,width=8cm,angle=0]{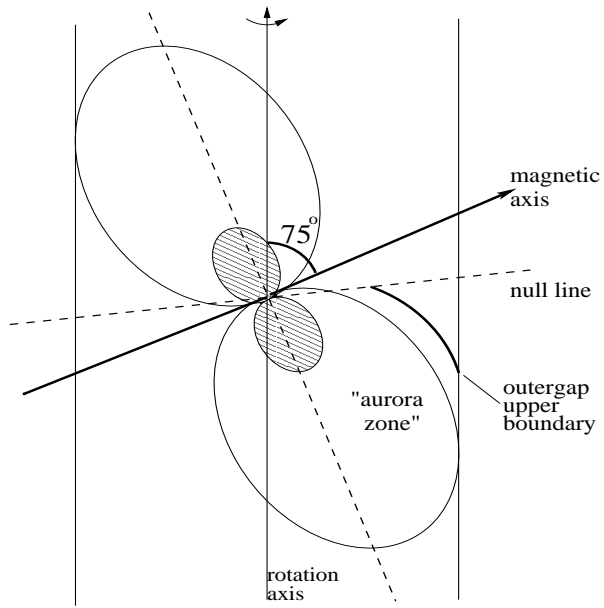}
\caption{\label{FigModel}A schematic picture of a physical model
for PSR~B1055--52, deduced from its radio and $\gamma$-ray
profiles. The corotation region, bounded by the fieldline $s=2$, is
shaded and lies well within region defined by the last closed
fieldline to touch the light cylinder.  It is surrounded by an
``auroral'' zone, which may or may not be entirely closed, but which
must support electric fields parallel to its magnetic fieldlines. It
makes two different intersections with the null surface, one very
close to the star and a second no longer as distant from the surface
as in a conventionally-defined corotation region. This may give rise
to a complex outergap structure, with pair-created particles
continually reflected between the null surfaces. The $\gamma$-ray
emission, appropriate to a ``middle-aged'' pulsar, is located above
the outergap, stretching from the null line to the light cylinder.}
\end{figure}

The kidney shaped central area of Fig. \ref{FigPolarMap}, which almost
completely encloses the footprints of the IP's line of sight,
highlights a significant consequence of the inferred geometry: the
active fieldlines of the IP are open fieldlines which not only do not
thread the outergap region beyond the null line close to the surface,
but they also do not thread the null surface at any point (assuming a
pure dipole geometry as in Fig. \ref{FigModel}) before they leave the
light cylinder. This suggests that along these fieldlines the pulsar
may expel or accrete charged particles without having to encounter any
charge sign boundary. Such fieldlines appear to bunch around the
magnetic pole bounded by the radial parameter $s=1$, the conventional
limit for a polar cap defined by the last closed fieldlines touching
the light cylinder. Thus the IP, if observed on a line of sight which
missed the MP, would not suggest an unusual geometry. 

By contrast, the fieldlines generating the MP have polar footprints
well beyond $s=1$ (Fig. \ref{FigPolarMap}), although the observed
emission region is itself well above the null line
(Fig. \ref{Fig1055field}). We are therefore led to the conclusion that
in PSR~B1055--52's MP we are seeing emission from regions of the
magnetosphere normally regarded as a closed and corotating. A polar
cap radius of $s=2$ is the footprint of a fieldline which closes at a
distance of only a quarter of the light cylinder radius and which
intersects the null line at a height of only 160 km above the polar
cap (Fig. \ref{FigModel}). In Fig. \ref{FigPolarMap} we see that radio
emission occurs on fieldlines with parameters $s=1.5$ (or
smaller) to $s=2$, indicating emission on fieldlines for a
significant region between the $s=2$ surface shaded in
Fig. \ref{FigModel} and the last closed fieldlines
($s\approx{1}$). Radio emission from these fieldlines implies particle
acceleration and hence the presence of electric fields parallel to the
fieldlines.  So it is conceivable that such fieldlines are open and
cross the light cylinder (as they do, for example, in the models of
aligned magnetospheres with reduced corotation regions given by
\citealt{gmm+04}). The closeness of the outergap to the neutron star
surface may lead to an active outergap accelerator merging with the
polar gap (see \citealt{hs02b,hir06,mh04,qlw+04}) and the much reduced
corotating region (shaded region in Fig. \ref{FigModel}).

Alternatively, the quasi-closed region between fieldlines $s=1$ and
$s=2$ can be seen as an auroral zone, analogous to those known to be
present in the planets, in which accretion from a debris disk may
occur and modulate the radio signal (\citealt{cs08,lm07}). In fact,
accretion could modulate the entire outergap region through an
injection of charge (\citealt{hs02b}).
Since all the surface $s=2$ is now so close to the neutron
star, it is not impossible to envisage interaction between the {\em
two} null surfaces which intersect it in the meridional plane
(Fig. \ref{FigModel}). Possibly ``gaps'' arise at both at which
pair-creation may be stimulated by thermal X-rays from the star's
surface leading to secondary particles ``bouncing'' between the gaps
and stimulating the observed MP emission. 

Whatever conditions are present in the envelope between $s=2$ and
$s=1$, PSR~B1055--52's $\gamma$-rays are almost certainly generated
relatively close to the light cylinder above an outer gap which
stretches from the last closed fieldline (whether $s=1$ or $s=2$)
towards the light cylinder. Unlike most other $\gamma$-ray pulsars,
B1055--52 has a single broad peak which is shown by \cite{wrwj08} to
match closely to the geometric predictions of the outergap model of
\cite{rom96a}, lying on a fieldline with $s=0.82$. Outergaps with
$\gamma$-ray emission regions lying significantly above the last
closed fieldline are thought to be a feature of relatively ``old''
pulsars, as PSR~B1055--52 indeed is by the standards of known
$\gamma$-ray pulsars. The fact that the $\gamma$-ray pulse overlaps
the phase of the MP indicates that both regions point to us at about
the same time, which lends support to our model (Fig. \ref{FigModel}).

Our conclusions about the geometry of PSR~B1055--52 result from a
rigid application of the RVM model to high-quality data, taking into
account the expected effects from aberration. We are convinced that
this pulsar emits radiation from fieldlines lying well outside the
conventionally-defined polar cap boundary. Consequently, we would
question the common assumption that all pulsars have closed,
quiescent, corotating regions stretching to the light cylinder.
Possibly this only occurs after the outergap has become inactive, so
that energetic (high $\dot{E}$) pulsars can often have wide radio
profiles, as observed \citep{wj08b}. Large polar caps might also help
explaining the inconsistency found in calculating emission heights of
energetic pulsars using last-closed-fieldline methods \citep{wj08b}.

An unsatisfactory aspect of our model is the one-sidedness of the MP
emission region. This might arise from the very different azimuthal
trajectories which particles on the trailing and leading halves must
take. On the trailing half they flow counter to the sense of the
pulsar's rotation and can smoothly cross the light cylinder to the
wave zone. However, those leaving the leading half of the polar cap
will increase speed and hence Lorentz factor as they gain height on
corotating fieldlines. It has been suggested that this may lead to
additional pair-creation and high-energy emission as the particles
free themselves from fieldlines near the light cylinder
(\citealt{mww76,dk82}), but subsequent screening and its consequences
for radio emission have not been considered.

\section*{Acknowledgments}
The authors would like to thank Simon Johnston, Don Melrose, Qinghuan
Luo, Matthew Verdon, Kyle Watters, Mike Keith and Wim Hermsen
for useful discussions, as well as the referee Jarek Dyks. GW thanks
the University of Sussex for an ongoing Visiting Research Fellowship,
and is grateful to the CSIRO for support for a stay at ATNF, during
which most of this work was carried out.  The Australia Telescope is
funded by the Commonwealth of Australia for operation as a National
Facility managed by the CSIRO.

\label{lastpage} 

\appendix
\section{Trigonometric relations}
\label{SctApp}

\begin{figure}
\begin{center}
\includegraphics[height=0.7\hsize,angle=0]{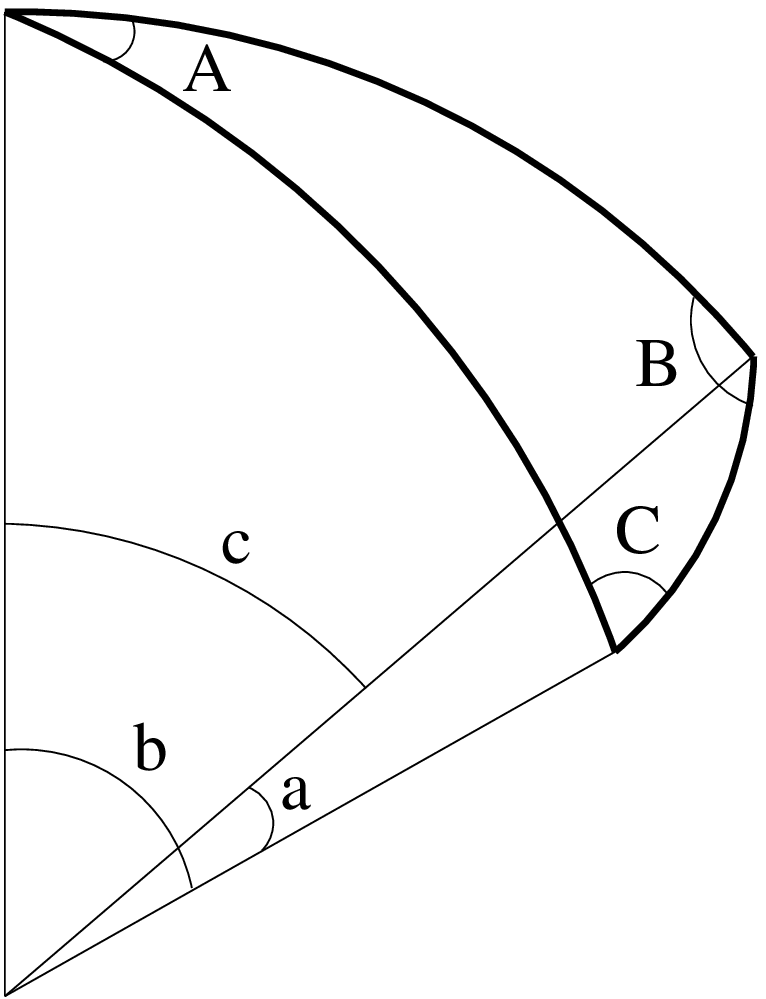}\\
\includegraphics[width=0.8\hsize,angle=0,trim=250 300 0 0, clip=true]{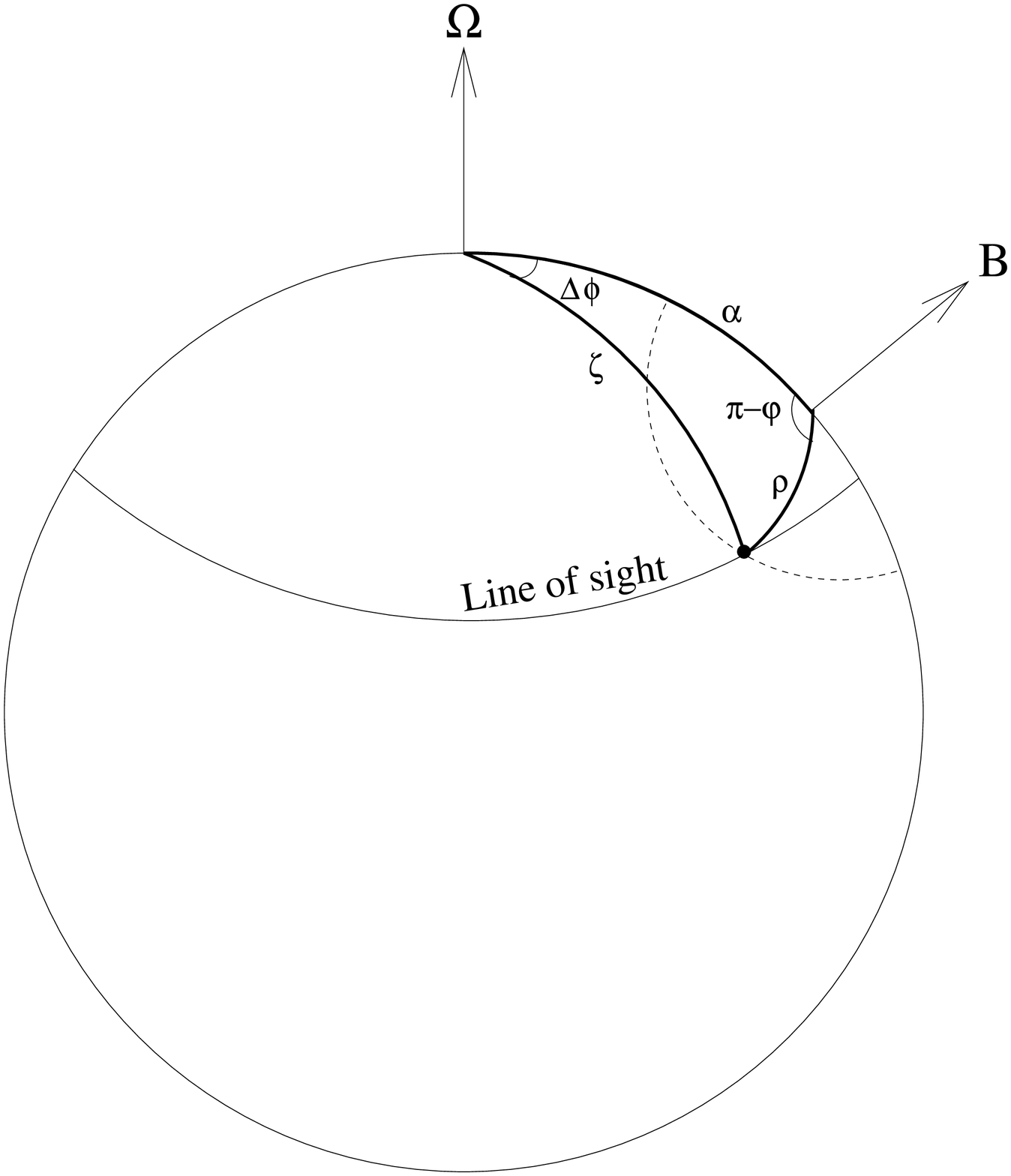}
\end{center}
\caption{\label{FigTrigonometry1}{\em Top plot:} A spherical
triangle. The angular lengths of the sides of the triangle are the
angles $a$, $b$ and $c$, while $A$, $B$ and $C$ are the angles between
the sides of the triangle measured on the surface of the sphere. {\em
Bottom plot:} A triangle on the celestial sphere used to derive the
half openings angle of the beam $\rho$ and the azimuthal angle of a
fieldline $\varphi$. The line of sight makes an angle $\zeta$ with the
rotation axis and and the magnetic axis makes an angle $\alpha$ with
the rotation axis.}
\end{figure}

We are interested to find the relation between the observed pulse
longitude and the magnetic coordinates $\rho$ (the angular distance
from the magnetic pole) and $\varphi$ (the azimuthal angle around the
magnetic pole). This coordinate transformation can be derived using
spherical geometry on the celestial sphere of the pulsar. Two
trigonometric relations which can be used are the sine and cosine
rules for sides (e.g. \citealt{sma60})
\begin{eqnarray}
\label{EqSin}
\frac{\sin a}{\sin A} &=& \frac{\sin b}{\sin B} = \frac{\sin c}{\sin C}\\
\label{EqCos}
\cos a&=&\cos b\cos c+\sin b\sin c\cos A.
\end{eqnarray}
See the top panel of Fig. \ref{FigTrigonometry1} for the definitions
of the angles in these equations.

The openings angle of the beam can be derived by applying the cosine
rule for sides (Eq. \ref{EqCos}) on the triangle plotted in the bottom
panel of Fig. \ref{FigTrigonometry1}:
\begin{eqnarray}
\cos \rho=\cos \alpha\cos \zeta+\sin \alpha\sin \zeta\cos \left(\Delta \phi\right)
\end{eqnarray}
(\citealt{ggr84}). Here
$\Delta \phi$ is the pulse longitude measured from the fiducial
plane. In the case of a profile which is symmetric around the fiducial
plane $\Delta \phi = W/2$.

The azimuthal angle $\varphi$ can be derived by using Eq. \ref{EqSin}
on the same triangle plotted in the bottom panel of
Fig. \ref{FigTrigonometry1} and one directly obtains
\begin{eqnarray}
\sin\varphi=\frac{\sin\zeta\sin\Delta\phi}{\sin\rho}
\end{eqnarray}
\citep{ggk+04}. It can be shown that this equation is equivalent to the following
expression which is used by \cite{wqxl06}:
\begin{eqnarray}
\label{EqDifficultVarphi}
\tan\left(\frac{\pi-\varphi}{2}\right)=\sqrt{\frac{\sin\left(p-\alpha\right)\sin\left(p-\rho\right)}{\sin p\sin\left(p-\zeta\right)}},
\end{eqnarray}
where $p = \frac{1}{2}\left(\alpha+\zeta+\rho\right)$.

\clearpage
\end{document}